\documentclass[%
superscriptaddress,
twocolumn,
showpacs,preprintnumbers,
amsmath,amssymb,
aps,
prl,
]{revtex4-1}

\usepackage{graphicx}
\usepackage{dcolumn}
\usepackage{bm}

\newcommand{\LSNROH}{LaSr$_{3}$NiRuO$_{4}$H$_{4}$}
\newcommand{\LSNRO}{LaSrNiRuO$_{4}$}
\newcommand{\uB}{\mu_{\rm B}}

\begin{document}


\title{Contrasting magnetism in isovalent layered LaSr$_{3}$NiRuO$_{4}$H$_{4}$ and LaSrNiRuO$_{4}$ due to distinct spin-orbital states}

\author{Xuan Wen}
 \affiliation{Laboratory for Computational Physical Sciences (MOE),
 State Key Laboratory of Surface Physics, and Department of Physics,
 Fudan University, Shanghai 200433, China}

\author{Ke Yang}
 \affiliation{Laboratory for Computational Physical Sciences (MOE),
 State Key Laboratory of Surface Physics, and Department of Physics,
 Fudan University, Shanghai 200433, China}


\author{Hua Wu}
\email{Corresponding author. wuh@fudan.edu.cn}
\affiliation{Laboratory for Computational Physical Sciences (MOE),
 State Key Laboratory of Surface Physics, and Department of Physics,
 Fudan University, Shanghai 200433, China}
\affiliation{Collaborative Innovation Center of Advanced Microstructures,
 Nanjing 210093, China}

%
%

\date{\today}

\begin{abstract}

The recently synthesized first $4d$ transition-metal oxide-hydride LaSr$_{3}$NiRuO$_{4}$H$_{4}$ with the unusual high H:O ratio surprisingly displays no magnetic order down to 1.8 K. This is in sharp contrast to the similar unusual low-valent Ni$^{+}$-Ru$^{2+}$ layered oxide LaSrNiRuO$_{4}$ which has a rather high ferromagnetic (FM) ordering Curie temperature $T_{\rm C}$ $\sim$ 250 K. In this work, using density functional calculations with aid of crystal field level diagrams and superexchange pictures, we find that the contrasting magnetism is due to the distinct spin-orbital states of the Ru$^{2+}$ ions (in addition to the common Ni$^{+}$ $S$=1/2 state but with a different orbital state): the Ru$^{2+}$ $S$=0 state in LaSr$_{3}$NiRuO$_{4}$H$_{4}$, but the Ru$^{2+}$ $S$=1 state in LaSrNiRuO$_{4}$. The Ru$^{2+}$ $S$=0 state has the $(xy)^2(xz,yz)^4$ occupation due to the RuH$_4$O$_2$ octahedral coordination, and then the nonmagnetic Ru$^{2+}$ ions dilute the $S$=1/2 Ni$^+$ sublattice which consequently has a very weak antiferromagnetic (AF) superexchange and thus accounts for no presence of magnetic order down to 1.8 K in LaSr$_{3}$NiRuO$_{4}$H$_{4}$. In strong contrast, the Ru$^{2+}$ $S$=1 state in LaSrNiRuO$_{4}$ has the $(3z^2-r^2)^2(xz,yz)^3(xy)^1$ occupation due to the planar square RuO$_4$ coordination, and then the multi-orbital FM superexchange between the $S$=1/2 Ni$^+$ and $S$=1 Ru$^{2+}$ ions gives rise to the high $T_{\rm C}$ in LaSrNiRuO$_{4}$. This work highlights the importance of spin-orbital states in determining the distinct magnetism.

\end{abstract}

\pacs{}
\maketitle


\section{I. Introduction}

Transition-metal (TM) oxides are a great platform for functional materials due to their diverse properties such as high temperature superconductivity, colossal magnetoresistance, and multiferroicity~\cite{Tokura2000,Dagotto2005,Hwang2012}. These abundant properties often stem from the intimate coupling of the charge, spin, and orbital degrees of freedom associated with electron correlations. In combination with doping, pressure, strain, and/or interfacial effects, which are able to tune the charge-spin-orbital states and to manipulate the materials properties, TM oxides are deemed a fertile field for exploration of new materials and novel properties~\cite{Tokura2000,Dagotto2005,Hwang2012,Ou2015,Fan2016}. Recently, anion doping is developed, along with the conventional cation doping, to modify the properties of TM oxides via changes of valence state, covalency, and band formation $etc$. In particular, substituting oxide by hydride is a typical anion doping which leads to emerging of TM oxide-hydride~\cite{Yajima2012,Hayward2002,Denis2014,Tassel2014}.

TM oxide-hydrides have attracted a lot of interest as they can produce an unusual low valence state, stronger covalency due to the lower electronegativity of hydrogen (compared with oxygen), and possibly enhanced band formation and magnetic coupling due to shortened TM-H-TM bonds. For example, $A$TiO$_3$ ($A$ = Ba, Sr, Ca) undergo an insulator-metal transition upon a hydride-for-oxide substitution into $A$TiO$_{3-x}$H$_y$~\cite{Yajima2012}. LaSrCoO$_3$H$_{0.7}$, SrVO$_2$H, and SrCrO$_2$H display high-temperature magnetic ordering~\cite{Hayward2002,Denis2014,Tassel2014}. Very recently, the first $4d$ TM oxide-hydride, {\LSNROH}, was synthesized from the Ruddlesden-Popper LaSr$_3$NiRuO$_8$ via topochemical anion exchange~\cite{Jin2018}. Its layered structure has the -(La,Sr)O-(Ni,Ru)H$_2$-(La,Sr)O- stacking sequence along the $c$-axis, see FIG.1(a). The Ni/Ru cations are bonded, along the $c$-axis, to two O anions of the neighboring (La,Sr)O sheets, and in the $ab$ plane, they are coordinated completely by the H anions after the topochemical anion exchange~\cite{Jin2018}. Thus, the H corner-shared (Ni,Ru)H$_4$O$_2$ octahedra form the (Ni,Ru)H$_2$ square planar sheets similar to the CuO$_2$ sheets in the superconducting cuprates.

{\LSNROH} does not display a long-range magnetic order down to 1.8 K and could thus well be paramagnetic (PM), according to the neutron diffraction and muon spin resonance measurements~\cite{Jin2018}. The absence of magnetic order was ascribed to the nonmagnetic Ru$^{2+}$ $S$=0 state, which dilutes the $S$=1/2 Ni$^+$ magnetic lattice and suppresses the superexchange interactions therein. This finding is in sharp contrast to the recent observation by the same group of authors that the isovalent layered {\LSNRO} [synthesized via a topochemical reduction, see FIG. 1(b)] has a FM order with a rather high $T_{\rm C}$ $\sim$ 250 K~\cite{Patino2016}. This FM order was suggested to be associated with the $S$=1/2 Ni$^+$ and $S$=0 Ru$^{2+}$ ground state~\cite{Patino2016}, which is the same as in {\LSNROH}.

Naturally, several questions arise: Why do the two similar layered materials in the `same' charge-spin state show such contrasting magnetism?
Does the picture of the $S$=1/2 Ni$^+$ and $S$=0 Ru$^{2+}$ ground state hold for both the materials? May the contrasting magnetism be due to distinct spin-orbital states (albeit the same Ni$^+$-Ru$^{2+}$ state)? To answer these questions, we have carried out a comparative study for {\LSNROH} and {\LSNRO}, by performing density functional theory (DFT) calculations and analyzing the crystal field level diagrams and superexchange interactions. Note that our previous study has shown that {\LSNRO} has the robust Ni$^{+}$ $S$=1/2 and Ru$^{2+}$ $S$=1 ground state~\cite{Zhu2017}, and here we provide more evidence from hybrid functional calculations to support this finding. Moreover, we have studied the hypothetic material LaSrNiZnO$_4$ with a substitution of the nonmagnetic Zn$^{2+}$ for the $S$=0 Ru$^{2+}$, to probe the magnetism of {\LSNRO} if the Ru$^{2+}$ is in the $S$=0 state. Indeed, the present work well explains the PM like behavior of {\LSNROH} and the strong FM of {\LSNRO}, by demonstrating that their contrasting magnetism is exactly due to the distinct spin-orbital states of the Ru$^{2+}$ ions and the subsequent very different superexchange interactions: the Ru$^{2+}$ $S$=0 state with the $(xy)^2(xz,yz)^4$ occupation in {\LSNROH}, but the Ru$^{2+}$ $S$=1 state with the $(3z^2-r^2)^2(xz,yz)^3(xy)^1$ occupation in {\LSNRO}. Thus, this work highlights the vital role of the combined charge-spin-orbital states in determining the distinct magnetism.

\section{II. Computational Details}

The DFT calculations were performed using the full-potential augmented plane
wave plus local orbital code (Wien2k)~\cite{blaha2001}. Using the experimental
lattice parameters of the Ni-Ru disordered {\LSNROH}, $a_0$=$b_0$=3.623~\AA~ and $c_0$=13.317~\AA~\cite{Jin2018},
we generate a Ni-Ru checkerboard ordered structure with $a$= $b$= $\sqrt{2}a_0$= 5.123~\AA~and $c$=$c_0$=13.317~\AA~for our calculations, see FIG. 1(a). The neglect of the Ni-Ru disorder does not affect the present discussion of the local crystal field and the very weak
magnetism of {\LSNROH}. The theoretically optimized lattice constants, $a$=$b$=5.157~\AA~ and $c$=13.405~\AA, are almost the same (within 1\%) as the experimental ones. And a 2$a_0\times$2$b_0\times$$c_0$ supercell is used to treat the intra-layer AF state of the Ni sublattice. The muffin-tin sphere radii
 of La, Sr, Ni, Ru, O, and H are chosen as 2.5, 2.5, 2.0, 2.0, 1.4, and 1.4
 Bohr, respectively. The plane-wave cut-off energy for interstitial wave
 functions is set to be 12 Ry, and a mesh of 5$\times$5$\times$3 k-points  was sampled for integration over the Brillouin zone.
The atomic relaxation is carried out using the local spin-density approximation (LSDA), till the atomic forces are each smaller than 25 meV/\AA. To describe the correlation effects of the Ni $3d$ and Ru $4d$ electrons, the LSDA plus Hubbard U (LSDA+U) method is employed~\cite{Anisimov1993},
with a common value of U=6 eV (3 eV) and $J_{\rm H}$=1 eV (0.6 eV) for the Ni $3d$ (Ru $4d$) states. To confirm our LSDA+U results, we have also performed the PBE+U calculations and the hybrid functional PBE0 calculations~\cite{Ernzerhof1999,Perdew1996,Tran2006}.
For {\LSNRO} and the hypothetic LaSrNiZnO$_4$ with Zn$^{2+}$ substitution for Ru$^{2+}$, we have used the experimental lattice parameters of {\LSNRO}, $a$=5.660~\AA, $b$=5.658~\AA, and $c$=6.901~\AA~\cite{Patino2016} (being almost the same as the optimized values, $a$=5.603~\AA, $b$=5.601~\AA, and $c$=6.831~\AA), and have used the same computational parameters as above (except for the k mesh, here 5$\times$5$\times$5).

 \begin{figure}[t]
\centerline{\includegraphics[width=7.5cm]{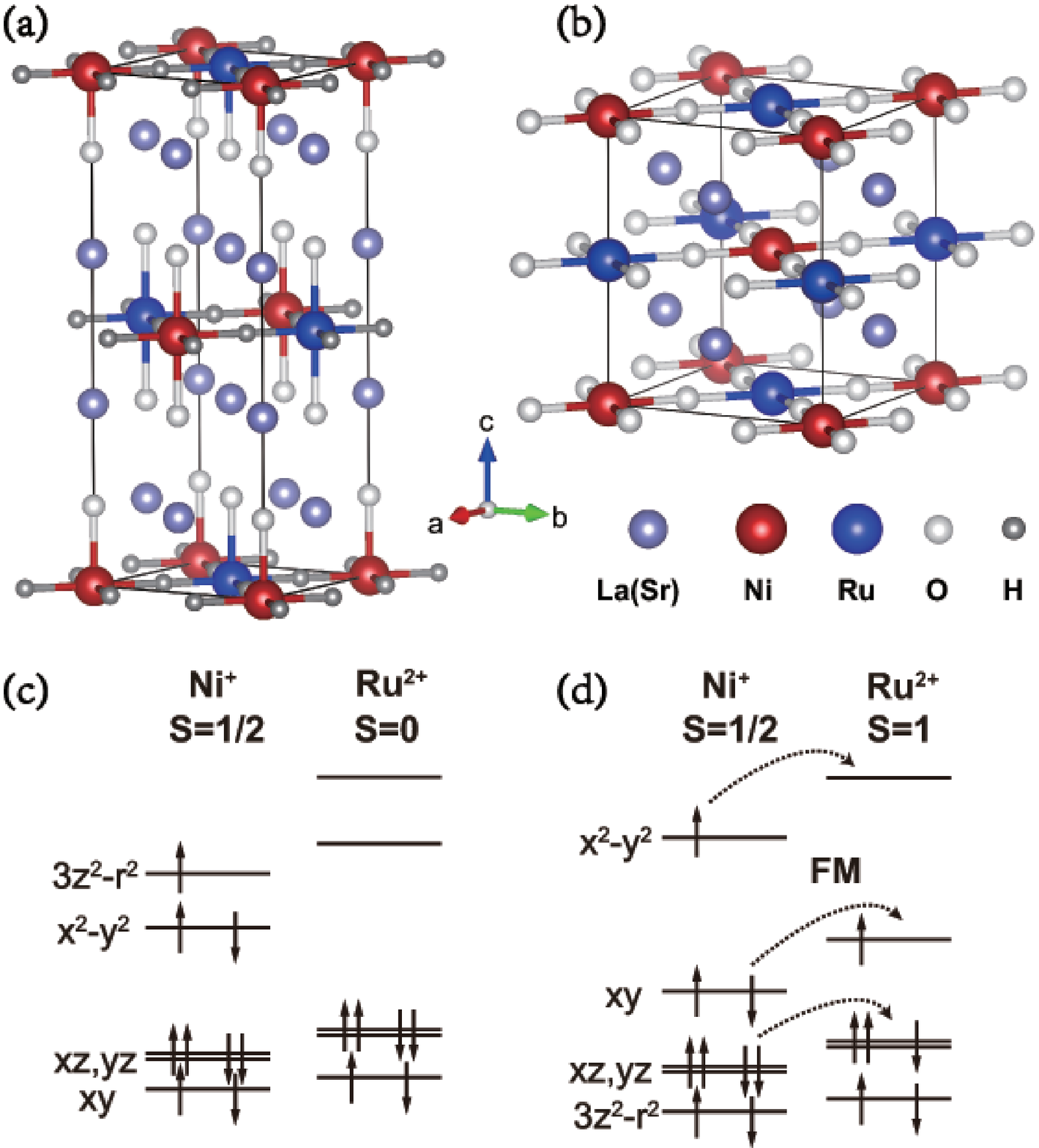}}
\caption{Crystal structure of (a) layered {\LSNROH} with the assumed Ni-Ru order (the local NiH$_4$O$_2$ and RuH$_4$O$_2$ octahedra), and of (b) {\LSNRO} with the local NiO$_4$ and RuO$_4$ planar squares. Schematic crystal field level diagrams of (c) Ni$^+$ $S$=1/2 and Ru$^{2+}$ $S$=0 in {\LSNROH}, and of (d) Ni$^+$ $S$=1/2 and Ru$^{2+}$ $S$=1 in {\LSNRO}. (c) A weak AF superexchange would occur in the Ni$^+$ $S$=1/2 sublattice of {\LSNROH}, and (d) the multi-orbital strong FM superexchange between Ni$^+$ $S$=1/2 and Ru$^{2+}$ $S$=1 would appear in {\LSNRO}.
}
\label{fig1:el}
\end{figure}

\section{III. Results and Discussion}

\subsection{Ni$^{+}$ $S$=1/2 and Ru$^{2+}$ $S$=0 state in {\LSNROH} and its very weak antiferromagnetism}

We first carry out LSDA calculations for {\LSNROH}. After atomic relaxation, the NiH$_4$O$_2$ octahedron has the planar Ni-H bondlengths of 1.821 ~\AA $\times$ 4 and the apical Ni-O bondlengths of 2.072 \AA$\times$2, and the RuH$_4$O$_2$ octahedron has the corresponding bondlengths of 1.802 \AA$\times$4 and 2.149 \AA$\times$2, see TABLE \uppercase\expandafter{\romannumeral1}. While the planar Ni-H (Ru-H) bondlength is very close to the experimental one of 1.811 \AA, the apical Ni-O (Ru-O) bondlength is smaller than the experimental one of 2.208 \AA~\cite{Jin2018} and this discrepancy could arise from the neglect of the actual atomic disorder of this layered material in the present calculations. Owing to the smaller H size than O, the planar Ni-H (Ru-H) bonds in {\LSNROH} are shorter than the planar Ni-O ones of 1.955 \AA$\times$4 on average (the Ru-O ones of 2.051 \AA$\times$4 on average) in {\LSNRO}~\cite{Patino2016}. However, the apical Ni-O and Ru-O bonds get longer in {\LSNROH} due to an release of the lattice strain in its layered structure.
\begin{table}[ht]
  \caption{The optimized and experimental bondlengths of {\LSNROH} (in unit of~\AA), the estimated $t_{2g}$-$e_g$ like crystal field splitting ($\Delta_{\rm CF}$, see FIG. 2) and Hund exchange energy $J_{\rm H}$ (both in unit of eV).
}
  \label{tb1}
  \begin{tabular}{c@{\hskip4mm}c@{\hskip4mm}c@{\hskip4mm}c@{\hskip4mm}c}
\hline\hline

  bond &      length  &  Expt~\cite{Jin2018} &   $\Delta_{\rm CF}$  & $J_{\rm H}$\\ \hline
  Ni-H     &   1.821$\times$4        &    1.811$\times$4& 1 & 1    \\
  Ni-O     &   2.072$\times$2        &    2.208$\times$2&   &         \\ \hline

  Ru-H     &   1.802$\times$4        &    1.811$\times$4& 2 & 0.6      \\
  Ru-O     &   2.149$\times$2        &    2.208$\times$2&   &           \\
\hline\hline
 \end{tabular}
\end{table}

Using the relaxed structure, the obtained FM solution has a total spin moment of 0.88 $\uB$ per formula unit (fu), which is carried mainly by the Ni ion (0.64 $\uB$), see TABLE \uppercase\expandafter{\romannumeral2}. The ligand O (H) ions each have the finite spin moment of 0.01 $\uB$ (0.01 $\uB$) due to the Ni-O (Ni-H) hybridizations, and the interstitial region per fu has the spin moment of 0.09 $\uB$. The Ru ion is weakly spin polarized (0.02 $\uB$), which is induced by the magnetic Ni sublattice. We now have a look at the calculated partial density of states (DOS) results, see FIG. 2. The Ni $3d$ orbitals are almost fully occupied, except for the minority-spin $3z^2-r^2$ orbital, being indicative of the Ni$^{+}$ ($3d^9$) $S$=1/2 state. The Ni $3d$ levels are ordered, from low to high, as $xy$, ($xz$,$yz$), $x^2-y^2$, and $3z^2-r^2$. Among the Ru $4d$ orbitals, the $t_{2g}$ type ($xy$, $xz$ and $yz$) orbitals are fully occupied, but the $e_g$ type ($x^2-y^2$ and $3z^2-r^2$) orbitals are fully unoccupied, thus showing the Ru$^{2+}$ ($4d^6$) $S$=0 state. While the Ru $4d$ has the same crystal field level sequence as the Ni $3d$, the magnitude of the crystal field splitting is different. For example, the Ru$^{2+}$ $t_{2g}$-$e_g$ energy splitting is about 2 eV due to the strong crystal field effect of the `fat' $4d$ orbitals, and it is much stronger than the Hund exchange of about 0.6 eV for Ru, thus stabilizing the nonmagnetic Ru$^{2+}$ $S$=0 ($t_{2g}^6$) state. In contrast, Ni$^{+}$ $3d$ orbitals have a reduced $t_{2g}$-$e_g$ splitting of about 1 eV.

\begin{table}[hb]
\begin{center}
  \caption{Relative total energies $\Delta$E (meV/fu), total and local spin moments ($\mu$$_{\rm B}$) of {\LSNROH}, {\LSNRO}, and hypothetic LaSrNiZnO$_4$ (with Zn$^{2+}$ substitution for Ru$^{2+}$) in the FM or intra-layer AF state given by LSDA, LSDA+U, and PBE0 calculations. The LSDA+U results of {\LSNRO} are adapted from Ref.~\cite{Zhu2017}. While {\LSNROH} is weakly AF coupled (in the Ni$^{+}$ $S$=1/2 and Ru$^{2+}$ $S$=0 ground state) and so is LaSrNiZnO$_4$ which models {\LSNRO} in the Ni$^{+}$ $S$=1/2 and Ru$^{2+}$ $S$=0 state, {\LSNRO} is in the strongly FM coupled Ni$^{+}$ $S$=1/2 and Ru$^{2+}$ $S$=1 ground state.
}
  \label{tb2}
  \begin{tabular}{l@{\hskip3mm}r@{\hskip3mm}c@{\hskip3mm}c@{\hskip3mm}c@{\hskip3mm}c}
\hline\hline
           &        & $\Delta$E & tot  & Ni   & Ru   \\ \hline
{\LSNROH}  & LSDA  ~FM  & -         & 0.88 & 0.64 & 0.02 \\ \hline
           &LSDA+U  ~FM  & 0         & 1.00 & 0.89 & 0.02 \\
           &      AF& --0.27     & 0.00 & 0.89 & 0.00 \\ \hline
    {\LSNRO}  &LSDA+U  ~FM  & 0         & 3.00 & 1.03 & 1.63      \\
           &   Ref.~\cite{Zhu2017}  ~AF& 126       & 0.00 & 0.87 & 1.48     \\ \hline
           &PBE0  ~FM  & 0         & 3.00 & 0.97 & 1.52     \\
           &      AF& 135       & 0.00 & 0.88 & 1.39     \\ \hline
    LaSrNiZnO$_4$  &LSDA+U  ~FM  & 0         & 0.92 & 0.98 & - \\
           &      AF& --2.36     & 0.00 & 1.05 & - \\
\hline\hline
 \end{tabular}
\end{center}
\end{table}

 \begin{figure}[b]
\centerline{\includegraphics[width=7cm]{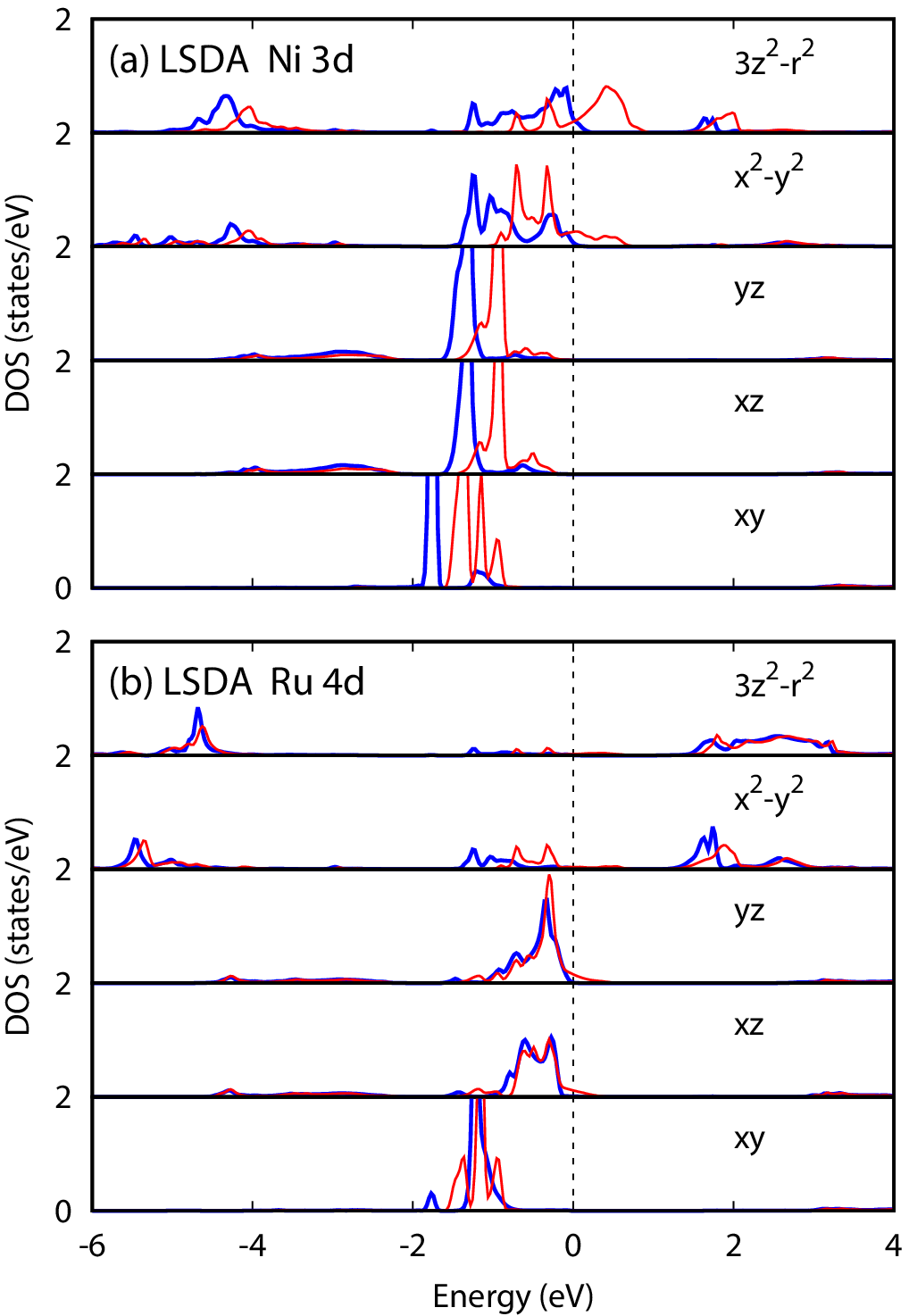}}
\caption{Ni $3d$ and Ru $4d$ DOS of {\LSNROH} calculated by LSDA. The blue (red) curves stand for the majority (minority) spin channel. The Fermi level is set at zero energy.
}
\label{fig2:el}
\end{figure}
 \begin{figure}[b]
\centerline{\includegraphics[width=7cm]{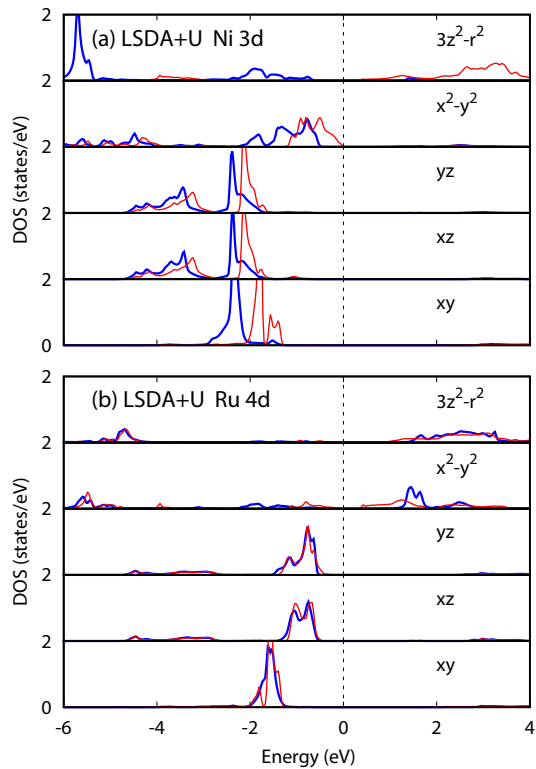}}
\caption{Ni $3d$ and Ru $4d$ DOS of {\LSNROH} calculated by LSDA+U. The blue (red) curves stand for the majority (minority) spin channel. The Fermi level is set at zero energy.
}
\label{fig2:el}
\end{figure}

Considering the formal H$^-$ and O$^{2-}$ charge states in the local NiH$_4$O$_2$ (RuH$_4$O$_2$) octahedron, and the in-plane Ni-H (Ru-H) bondlength of 1.821 \AA~ (1.802 \AA) and the out-of-plane Ni-O (Ru-O) bondlength of 2.072 \AA~ (2.149 \AA), and using a simple point charge model of crystal field Coulomb potential [being proportional to $Ne/r$ where $N$ is the charge state (--1 for H$^{-}$ and --2 for O$^{2-}$) and $r$ is a bondlength], one can establish the crystal field level diagrams. It is easily understood that in the local octahedral NiH$_4$O$_2$ and RuH$_4$O$_2$ surrounding, the in-plane $xy$ orbital feels the weakest crystal field Coulomb repulsion and the out-of-plane $3z^2-r^2$ sees the strongest one. Therefore, both the Ni $3d$ and Ru $4d$ crystal field levels have the energy sequence, from low to high, $xy$, ($xz$,$yz$), $x^2-y^2$, and $3z^2-r^2$, see FIG. 1(c). Indeed, this level sequence is confirmed by the above LSDA calculations, see the DOS results in FIG. 2. As a result, {\LSNROH} has the formal Ni$^+$ $S$=1/2 ($3d^9$) state with the electronic configuration ($xy$)$^2$($xz$,$yz$)$^4$($x^2-y^2$)$^2$($3z^2-r^2$)$^1$, and the Ru$^{2+}$ $S$=0 ($4d^6$) state with ($xy$)$^2$($xz$,$yz$)$^4$($x^2-y^2$)$^0$($3z^2-r^2$)$^0$. These formal valence-spin-orbital states well account for the above calculated spin moments reduced by the Ni-H and Ni-O covalency.

Owing to the common overestimation of electron delocalization by LSDA, the broadened Ni$^{+}$ $3z^2-r^2$ and $x^2-y^2$ bands stride over the Fermi level and form a metallic solution, see FIG. 2. In order to reproduce the insulating behavior of {\LSNROH}, electron correlation effects should be included. To do so, we have carried out LSDA+U calculations, which include the orbital-dependent Coulomb interactions for the Ni $3d$ and Ru $4d$ states. The obtained insulating solution is shown in FIG. 3. Now the Ni$^{+}$ $S$=1/2 and Ru$^{2+}$ $S$=0 valence-spin-orbital states are more clear, and only the half filled Ni$^{+}$ $3z^2-r^2$ orbital is magnetically active and contributes to the sole $S$=1/2. The enhanced electron localization results in an increasing Ni$^{+}$ spin moment from 0.64 $\uB$ by LSDA to 0.89 $\uB$ by LSDA+U, see TABLE \uppercase\expandafter{\romannumeral2}.

Note that the $xy$ and ($xz$,$yz$) levels of the Ni$^{+}$ $3d$ orbitals seem to interchange their sequence by a comparison between FIGs. 3(a) and 2(a). This is because the minority-spin Ni$^{+}$ $3d$ $e_g$ orbital occupations get more distinct (all other $3d$ orbitals being fully occupied), from ($x^2-y^2$)$_\downarrow^{0.69}$($3z^2-r^2$)$_\downarrow^{0.39}$ by LSDA to ($x^2-y^2$)$_\downarrow^{0.82}$($3z^2-r^2$)$_\downarrow^{0.12}$ by LSDA+U. Then, the enhanced anisotropic inter-orbital interaction between the planar $x^2-y^2$ and $xy$ orbitals raises the $xy$ level [compared with the interaction between the out-of-plane $3z^2-r^2$ and ($xz$,$yz$) orbitals], making the otherwise lowest $xy$ crystal field level higher than the ($xz$,$yz$) levels, see FIG. 3(a). In addition, the singly occupied Ni$^{+}$ $3z^2-r^2$ level gets lowest due to the absence of the intra-orbital Coulomb repulsion, compared with other four doubly occupied Ni$^{+}$ $3d$ orbitals. In contrast, owing to the closed $t_{2g}^6$ ($S$=0) subshell, the Ru$^{2+}$ $4d$ crystal field level sequence remains unchanged, by a comparison between FIGs. 3(b)  and 2(b). In principle, it is inappropriate to use DFT+U calculations including orbital-dependent Coulomb interactions to draw a single-electron crystal field level diagram as done in Ref.~\cite{Jin2018} [see its FIG. 1(b)], where the lowest ($xz$,$yz$) and the highest $3z^2-r^2$ are self-contradicting.

The insulating {\LSNROH} has the $S$=1/2 Ni$^{+}$ and $S$=0 Ru$^{2+}$ state. Therefore, the magnetic Ni$^{+}$ sublattice is diluted by the nonmagnetic Ru$^{2+}$ ions and then would have a weak AF superexchange interaction (if any) via the Ni$^{+}$ $3z^2-r^2$ orbital. This anticipation is indeed confirmed by our LSDA+U calculations which show that the intra-layer AF state of this layered material is slightly more favorable than the FM state by 0.27 meV/fu, see TABLE \uppercase\expandafter{\romannumeral2}. Note that our PBE+U calculations give almost the same results to the LSDA+U ones, with the corresponding FM-AF energy difference of 0.14 meV/fu and the Ni$^{+}$ spin moment of 0.91 $\mu_{\rm B}$ (the same as in Ref.~\cite{Jin2018}) for the $S$=1/2 Ni$^{+}$ and $S$=0 Ru$^{2+}$ ground state. Thus, these results well account for the experimental observation that {\LSNROH} is not magnetically ordered down to 1.8 K and it could well be PM~\cite{Jin2018}.

The PM behavior of {\LSNROH} is exactly due to the dilution of the magnetic $S$=1/2 Ni$^{+}$ sublattice by the nonmagnetic $S$=0 Ru$^{2+}$ ions. Note that this PM state is not a nonmagnetic state, and if we assume both the nonmagnetic $S$=0 Ni$^{+}$-Ru$^{2+}$ ions, the calculated total energy turns out to be higher than the $S$=1/2 Ni$^{+}$ and $S$=0 Ru$^{2+}$ ground state by 27 meV/Ni in LSDA and by 1510 meV/Ni in LSDA+U. Moreover, this $S$=1/2 Ni$^{+}$ and $S$=0 Ru$^{2+}$ ground state is further verified by the fixed-spin-moment calculation and hybrid functional PBE0 calculation. As the FM and AF states are almost degenerate, for simplicity the FM state is used here. Our fixed-spin-moment calculation within LSDA+U, assuming the $S$=1 Ru$^{2+}$ and $S$=1/2 Ni$^{+}$ state, shows that this hypothetic state is much higher in energy than the ground state by 1397 meV/fu. This energy value just matches the spin-state excitation of the Ru$^{2+}$ ion from $S$=0 to $S$=1, at the cost of the $t_{2g}$-$e_g$ crystal field excitation [about 2 eV, see FIG. 2(b)] but with the gain of Hund exchange (about 0.6 eV for Ru). These calculations show that the Ru$^{2+}$ $S$=0 ground state (together with Ni$^{+}$ $S$=1/2) is robust and the Ru$^{2+}$ $S$=1 state would be highly unstable. Moreover, our hybrid functional PBE0 calculation again confirms the stable Ni$^{+}$ $S$=1/2 and Ru$^{2+}$ $S$=0 ground state, giving the local spin moments of 0.83 $\uB$/Ni$^{+}$ and 0.05 $\uB$/Ru$^{2+}$, being well comparable with the above LSDA+U results of 0.89 $\uB$/Ni$^{+}$ and 0.02 $\uB$/Ru$^{2+}$. Therefore, all these calculations consistently arrive at the Ni$^{+}$ $S$=1/2 and Ru$^{2+}$ $S$=0 ground state, which well accounts for the PM behavior of {\LSNROH}. Note that all the above calculations and discussion are based on the Ni-Ru ordered structure, and that the actual Ni-Ru disorder would further suppress the weak magnetic coupling, thus giving rise to the PM behavior.

\subsection{{\LSNRO}: Strong FM with Ru$^{2+}$ $S$=1 $versus$ weak AF with Ru$^{2+}$ $S$=0}

The Ni$^{+}$ $S$=1/2 and Ru$^{2+}$ $S$=0 ground state of {\LSNROH} and its PM behavior prompt us to think about whether the same charge-spin ground state suggested for the layered {\LSNRO} would determine the rather high FM ordering temperature of about 250 K in {\LSNRO}~\cite{Patino2016}.
{\LSNRO} has the planar NiO$_2$-RuO$_2$ square sheet, and then the local NiO$_4$ and RuO$_4$ square coordinations without apical oxygens generate the tetragonal crystal field level sequence, from low to high, $3z^2-r^2$, ($xz$,$yz$), $xy$, and $x^2-y^2$ for Ni $3d$ and Ru $4d$ orbitals, see FIGs. 1(b) and 1(d). The Ni$^{+}$ has no other configuration than ($3z^2-r^2$)$^2$($xz$,$yz$)$^4$($xy$)$^2$($x^2-y^2$)$^1$ with $S$=1/2. The Ru$^{2+}$ is either in the $S$=1 state with the configuration ($3z^2-r^2$)$^2$($xz$,$yz$)$^3$($xy$)$^1$($x^2-y^2$)$^0$ or in the $S$=0 state with ($3z^2-r^2$)$^2$($xz$,$yz$)$^4$($xy$)$^0$($x^2-y^2$)$^0$, as the $S$=2 state with ($3z^2-r^2$)$^2$($xz$,$yz$)$^2$($xy$)$^1$($x^2-y^2$)$^1$ can simply be excluded due to the too high $x^2-y^2$ level in the local RuO$_4$ square. Then the Ru$^{2+}$ $S$=1 state competes with the $S$=0 state, depending on the interplay of Hund exchange and crystal field splitting between ($xz$,$yz$) and $xy$. While Ru $4d$ Hund exchange is about 0.6 eV, the $t_{2g}$-like crystal field splitting is normally small (being few tens or hundreds of meV~\cite{Wu2006}, here about 0.2 eV according to our previous LSDA calculation~\cite{Zhu2017}). Therefore, the Hund exchange dominates over the crystal field and favors the $S$=1 state, see FIG. 1(d).

As the Ru$^{2+}$ $S$=0 state cannot be stabilized in our LSDA+U calculations, the magnetism of {\LSNRO} in the Ni$^{+}$ $S$=1/2 and Ru$^{2+}$ $S$=0 state may not be directly probed in the present work. Therefore, we attempt to achieve it in a compromise way, by studying the hypothetic material LaSrNiZnO$_4$ with a substitution of the nonmagnetic Zn$^{2+}$ for the $S$=0 Ru$^{2+}$. This choice is also justified by the consideration that the Zn$^{2+}$ ion (0.74 \AA) has a very similar ionic size to the $S$=0 Ru$^{2+}$ ($t_{2g}^6$, probably 0.74 \AA) which is unavailable but can be extrapolated from the ionic sizes of 0.565 \AA/Ru$^{5+}$ ($t_{2g}^3$), 0.62 \AA/Ru$^{4+}$ ($t_{2g}^4$), and 0.68 \AA/Ru$^{3+}$ ($t_{2g}^5$) with a gradual electron filling of the $t_{2g}$ shell~\cite{Shannon1976}. As seen in TABLE \uppercase\expandafter{\romannumeral2}, the intra-layer AF state of LaSrNiZnO$_4$ turns out to be more favorable than the FM state only by 2.36 meV/fu. This weak AF interaction arises from the superexchange of the half-filled $x^2-y^2$ orbital of the $S$=1/2 Ni$^{+}$ ions via the nonmagnetic O$^{2-}$ ion and the Zn$^{2+}$ ion standing for the nonmagnetic $S$=0 Ru$^{2+}$ ion. It is not surprising that this AF superexchange in LaSrNiZnO$_4$ due to the in-plane Ni$^{+}$ $x^2-y^2$ orbital is `stronger' than the AF superexchange in {\LSNROH} with the out-of-plane Ni$^{+}$ $3z^2-r^2$ orbital, as the $x^2-y^2$ orbital has a larger overlap with the in-plane ligands. It is now clear that the Ni$^{+}$ $S$=1/2 and Ru$^{2+}$ $S$=0 state can only maintain a very weak AF in {\LSNRO} as in {\LSNROH}, and that this charge-spin state cannot at all yield the actual strong FM in {\LSNRO}.

 \begin{figure}[t]
\centerline{\includegraphics[width=7cm]{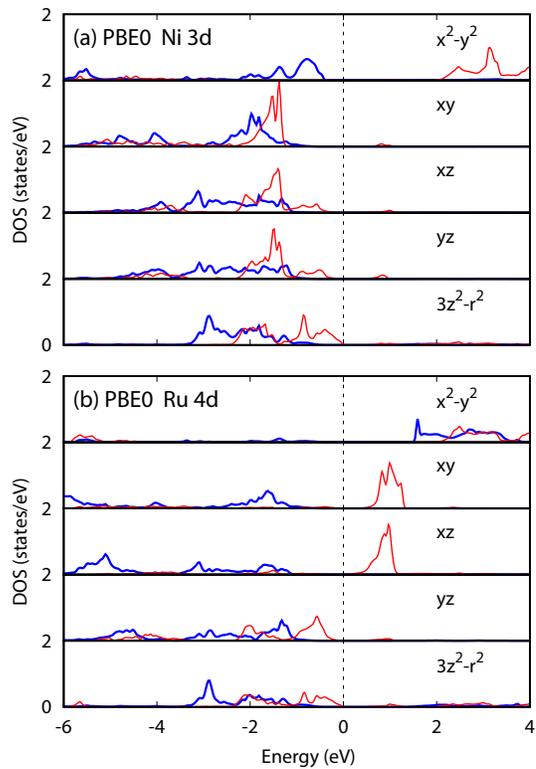}}
\caption{Ni $3d$ and Ru $4d$ DOS of {\LSNRO} calculated by PBE0. The blue (red) curves stand for the majority (minority) spin channel. The Fermi level is set at zero energy.
}
\label{fig2:el}
\end{figure}

Our previous LSDA+U calculations have shown that {\LSNRO} is in the robust Ni$^{+}$ $S$=1/2 and Ru$^{2+}$ $S$=1 ground state~\cite{Zhu2017}. Here again, we perform hybrid functional PBE0 calculations to verify this finding. The obtained results are quite similar to our previous LSDA+U results, see TABLE \uppercase\expandafter{\romannumeral2}. Moreover, we plot in FIG. 4 the partial DOS results of the insulating {\LSNRO} given by PBE0. The unique charge-spin-orbital states---Ni$^{+}$ $S$=1/2 and Ru$^{2+}$ $S$=1 are evident. In particular, the virtual hoppings of the majority-spin $x^2-y^2$ electron and the minority-spin $xy$ and ($xz$,$yz$) electrons between the $S$=1/2 Ni$^{+}$ and $S$=1 Ru$^{2+}$ ions produce the multi-orbital strong intra-layer (and considerably large inter-layer) FM superexchange interactions, as sketched in FIG. 1(d). Indeed, the intra-layer AF state turns out by our PBE0 calculations to be much higher in energy than the FM ground state by 135 meV/fu, which is close to the corresponding value of 126 meV/fu given by our previous LSDA+U calculations, see TABLE \uppercase\expandafter{\romannumeral2}. Then, all the above results prove that the rather strong FM in {\LSNRO} can not be explained by the $S$=0 Ru$^{2+}$ and $S$=1/2 Ni$^{+}$ state as suggested in another study~\cite{Patino2016}, but well be explained by the robust $S$=1 Ru$^{2+}$ and $S$=1/2 Ni$^{+}$ ground state.
\section{IV. Summary}

We have carried out a comparative study on two new isovalent Ni$^{+}$-Ru$^{2+}$ layered materials {\LSNROH} and {\LSNRO} to understand their contrasting magnetism, using density functional calculations, crystal-field level diagrams, and superexchange pictures. Our results show that the local NiH$_4$O$_2$ and RuH$_4$O$_2$ octahedral coordination in the former and the local NiO$_4$ and RuO$_4$ square coordination in the latter yield different crystal field level sequences and different energy splittings. Then their distinct spin-orbital states give rise to their contrasting magnetism: The Ni$^{+}$ $S$=1/2 state with the electronic configuration ($xy$)$^2$($xz$,$yz$)$^4$($x^2-y^2$)$^2$($3z^2-r^2$)$^1$ and the nonmagnetic Ru$^{2+}$ $S$=0 state with ($xy$)$^2$($xz$,$yz$)$^4$($x^2-y^2$)$^0$($3z^2-r^2$)$^0$ are responsible for the very weak antiferromagnetism of {\LSNROH} or even paramagnetism. In sharp contrast, the Ni$^{+}$ $S$=1/2 state with ($3z^2-r^2$)$^2$($xz$,$yz$)$^4$($xy$)$^2$($x^2-y^2$)$^1$ and the Ru$^{2+}$ $S$=1 state with ($3z^2-r^2$)$^2$($xz$,$yz$)$^3$($xy$)$^1$($x^2-y^2$)$^0$ determine the rather strong ferromagnetism in {\LSNRO} via the multi-orbital superexchange. This work highlights the vital role of particular spin-orbital states in determining distinct magnetism of transition-metal compounds.

\section{Acknowledgements}

This work was supported by the NSF of China (Grants No. 11674064 and No. 11474059) and by the National Key Research and Development Program of China (Grant No. 2016YFA0300700).

\nocite{*}

\bibliography{LSNROH}
\end{document}